\documentclass[a4paper,11pt]{article}

\usepackage{pos}
\usepackage{amsmath}
\usepackage{caption}
\usepackage{braket}
\usepackage{slashed}
\usepackage{bbold}
\usepackage{graphicx}
\usepackage {graphpap}
\usepackage{tikz}
\usepackage{fontenc}
\usepackage{times}
\usepackage{mathptmx}
\usepackage[utf8]{inputenc}

\usepackage{subcaption}

\title{Implementation of Simultaneous Inversion of a Multi-shifted Dirac Matrix for Twisted-Mass Fermions within DD$\alpha$AMG}
\ShortTitle{Simultaneous Inversion within DD$\alpha$AMG}

\author*[a,1]{Shuhei Yamamoto}
\author[a,1]{Simone Bacchio}
\author[a,1]{Jacob Finkenrath}

\affiliation[a]{Computation-based Science and Technology Research Center, The Cyprus Institute, \\
  Aglantzia, Nicosia, Cyprus}
\note{PRACE Collaboration}

\emailAdd{s.yamamoto@cyi.ac.cy}

\abstract{At physical light quark masses, efficient linear solvers are crucial for carrying out the millions of inversions of the Dirac matrix required for obtaining high statistics in quark correlation functions. Adaptive algebraic multi-grid methods have proven to be very efficient in such cases, exhibiting mild critical slowing down towards very light quark masses and outperforming traditional solver methods, such as the conjugate gradient method, at the physical point.   
We will discuss our implementations of simultaneous inversion of a (degenerate) Dirac matrix for twisted-mass fermions for multiple right-hand-sides (rhs) with multi-shifts and block-Krylov solvers. The implementation is carried out within the community library DD$\alpha$AMG, which implements aggregation-based Domain Decomposition adaptive algebraic multi-grid methods.  The block-Krylov solvers are provided via the Fast Accurate Block Linear krylOv Solver (Fabulous) library and can be used at coarser levels. 
Our code inverts Dirac matrices with different twisted-mass terms and for multiple rhs simultaneously and is thus also suitable for components within a typical lattice QCD simulation workflow, such as the rational approximation.  We  show preliminary results on scalability and compare the performance of our implementation when using different Block-Krylov solver techniques.}

\FullConference{
 The 38th International Symposium on Lattice Field Theory, LATTICE2021
  26th-30th July, 2021
  Zoom/Gather@Massachusetts Institute of Technology
}

\begin{document}
\maketitle
\section{Introduction}
\label{Sec:Intro}
In lattice QCD, inversion of Dirac matrices for a given right hand side (rhs) is a major computational task, performed repeatedly many times, for example, in measurement of observables and generation of gauge configurations.  Due to its importance, a number of algorithms to efficiently solve linear systems are proposed.  These algorithms make use of the nature of the matrix to be inverted, i.e., being large and sparse.  Typically in lattice QCD, approaches based on Krylov-subspace methods are adopted.  These methods are projection-based where the matrix is projected onto a subspace, called Krylov subspace, which is spanned by vectors obtained by repeated application of the matrix onto an initial guess, $x_0$.  They work well for large and sparse matrices such as Dirac matrices.  In particular, the conjugate gradient (CG) method has been used extensively.
However, the CG method is known to suffer from critical slowing down when the quark mass becomes smaller, resulting in the exponential increase in convergence time.  

To alleviate this issue, various methods have been tested, and it turned out that a class of solvers based on multigrid preconditioning is effective in reducing convergence time \cite{Luscher:2007se,Luscher:2007es,Osborn:2010mb,Babich:2010qb,Frommer:2013kla}.  There are various implementations based on multigrid approaches.  For clover-Wilson fermions, for example, Ref.~\cite{Luscher:2007se} proposes a two-level multigrid approach based on L\"{u}scher's inexact deflation.
Alternatively, multigrid approaches are applied to the generalized conjugate residual (MG-GCR) method in \cite{Brannick:2007ue,Clark:2008nh,Babich:2009pc,Babich:2010qb}.
Here, we will focus on an aggregation-based domain-decomposition multigrid approach (DD$\alpha$AMG), first proposed in Ref.~\cite{Frommer:2013kla} , adapted to degenerate twisted-mass fermions in Ref.~\cite{Alexandrou:2016izb}, and generalized to non-degenerate twisted-mass fermions in Ref.~\cite{Alexandrou:2018wiv}.
For the sake of completeness, multigrid approaches are adopted also for other fermions.  For instance, MG-GCR is extended for the case of domain-wall fermions \cite{Cohen:2012sh}, and DD$\alpha$AMG for overlap fermions \cite{Brannick:2014vda}. 

In what follows, we will discuss added capabilities of simultaneous inversion of multiple rhs and the usage of block-Krylov solvers in the DD$\alpha$AMG software package.

\section{DD$\alpha$AMG}
\label{Sec:DDalphaAMG}
In this section, we briefly review the basics of DD$\alpha$AMG.  As is mentioned in Section \ref{Sec:Intro}, DD$\alpha$AMG has been proposed to overcome the difficulties encountered by traditional Krylov-subspace methods such as CG.  For this purpose, it incorporates two preconditioners, namely smoother and coarse-grid correction.  
The smoother acts on higher modes of functions on the lattice, and coarse-grid correction works on lower modes with the help of restriction and prolongation operators.  The restriction operator extracts low-mode information and project onto a coarse lattice.  Then, the prolongation operator interpolates from the coarse lattice to the original fine lattice.  Various preconditioning procedures can be combined with coarse-grid correction at each coarse level.
In this work, we have adopted the red-black Schwarz alternating procedure (SAP) \cite{Luscher:2007es} for the smoother, and for the coarse-grid correction, we have chosen to use algebraic multigrid (AMG) \cite{10.5555/870015}.  
In application of SAP, the given lattice is divided into small blocks in a chessboard manner, and inversions of the local Dirac matrices, which are defined on the blocks, are performed.  So SAP removes UV-modes from the error after its application \cite{Luscher:2007es}.
As such, it suppresses the error in the higher end of the spectrum.

On the other hand, AMG focuses on the low modes of the lattice.  It is formulated as follows.  First, we define two operators, the prolongation operator, $P$, and the restriction operator, $R$.  The operators, $P$ and $R$, are defined algebraically using a few approximate low-mode eigenvectors based on the idea of L\"{u}scher's inexact deflation \cite{Luscher:2007se}.  Due to local coherence, this construction captures low-mode characteristics of the lattice \cite{Luscher:2007se}.  Now, with the help of these operators, one can define a coarse lattice and a coarse-grid operator.  The solution on the coarse grid is then prolonged back to the original lattice to produce an estimate.
This construction can be extended to multiple levels by applying it recursively on the coarse lattices.
Application of coarse-grid correction reduces the relative error in low modes.  So if we combine the two preconditioners, the SAP and coarse-grid correction, we can suppress the relative error in both high and low end of the eigenspectrum.

A bottleneck of DD$\alpha$AMG is its strong scalability at fix volume with increasing level of parallelization.  
\begin{figure}[hbt]
\centering
\includegraphics[width=0.45\textwidth]{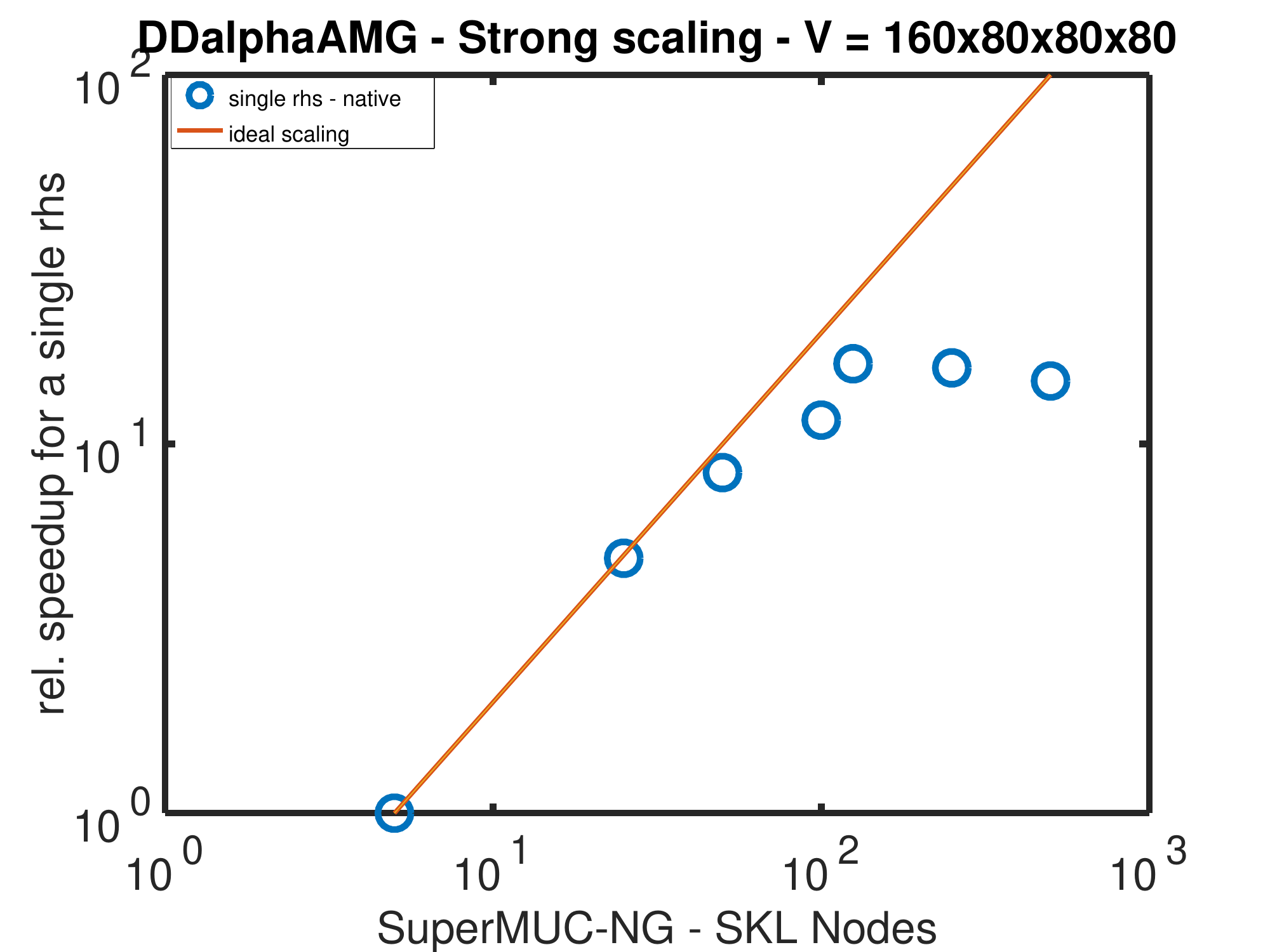}
\caption{A scaling plot on the ensemble of $N_f=2+1+1$ twisted mass clover with $a \sim 0.07$fm and $V=80^3\times160$ at physical point simulated on SuperMUC-NG (Intel Xeon ("Skylake")) at The Leibniz Supercomputing Center.  The orange solid line is the plot of ideal scaling, while the blue dots indicate the actual scaling behavior.}
\label{Fig:MPIscaling}
\end{figure}
Figure \ref{Fig:MPIscaling} depicts the scaling behavior on SuperMUC-NG equipped with Intel Xenon "Skylake" at The Leibniz Supercomputing Center.
For higher node counts, the speed-up diverges from the ideal scaling, and stagnation of performance beyond 125 Skylake nodes is observed for a three-level MG approach.  Given the current hardware trend, core counts per node will increase, and this means that the window for ideal scaling will shrink even further.  To overcome this issue, we have implemented simultaneous inversion of multiple rhs, which we will discuss in the next section.

\section{Multiple Right Hand Sides}
\label{Sec:MRHS}
Prior to implementation of multiple rhs, the original code at \url{https://github.com/sbacchio/DDalphaAMG} inverted each rhs one by one when it is given more than one rhs.  To gain benefit from SIMD architecture, loops over lattice objects such as gauge fields and lattice vectors are vectorized explicitly.  This vectorization was done by manually rewriting these loops with instruction sets for a specific SIMD extension.  However, such explicit vecotrization brings inconvenience as we need to rewrite all vectorized loops for a different SIMD extension with its own instruction sets.  Hence, in the version available at \url{https://github.com/sy3394/DDalphaAMG}, we decided to use the autovectorization feature of compilers by unrolling loops over multiple rhs.  In this way, we let compilers perform optimization analysis and vectorization.  This change has improved portability of our code and simplified its maintenance.

\begin{figure}[hbt] 
\centering
\begin{tikzpicture}
\node[rectangle, style={draw, thick, fill=red!20, minimum width=1.4cm,minimum height=0.6cm}] (1) at (0,0) {$v_0$};
\node[rectangle, style={draw, thick, fill=blue!20, minimum width=1.4cm,minimum height=0.6cm}] (2) at (1.4,0) {$v_1$};
\node[rectangle, style={draw, thick, fill=green!20, minimum width=1.4cm,minimum height=0.6cm}] (3) at (2.8,0) {$v_2$};
\node[rectangle,style={draw, thick, fill=red!20,minimum width=0.6cm,minimum height=0.6cm}] (5) at (5,0) {$v_0$};
\node[rectangle,style={draw, thick, fill=blue!20,minimum width=0.6cm,minimum height=0.6cm}] (6) at (5.6,0) {$v_1$};
\node[rectangle,style={draw, thick, fill=green!20,minimum width=0.6cm,minimum height=0.6cm}] (7) at (6.2,0) {$v_2$};
\node[rectangle,style={draw, thick, fill=red!20,minimum width=0.6cm,minimum height=0.6cm}] (8) at (6.8,0) {$v_0$};
\node[rectangle,style={draw, thick, fill=blue!20,minimum width=0.6cm,minimum height=0.6cm}] (9) at (7.4,0) {$v_1$};
\node[rectangle,style={draw, thick, fill=green!20,minimum width=0.6cm,minimum height=0.6cm}] (10) at (8,0) {$v_2$};
\node[rectangle,style={draw, thick, fill=red!20,minimum width=0.6cm,minimum height=0.6cm}] (11) at (8.6,0) {$v_0$};
\node[rectangle,style={draw, thick, fill=blue!20,minimum width=0.6cm,minimum height=0.6cm}] (12) at (9.2,0) {$v_1$};
\node[rectangle,style={draw, thick, fill=green!20,minimum width=0.6cm,minimum height=0.6cm}] (13) at (9.8,0) {$v_2$};
\draw[->,style={thick}] (3) -- (5);
\end{tikzpicture}
\caption{The comparison of how vectors are organized before and after implementation of simultaneous inversion of multiple r.h.s. The left figure show vector ordering in the old implementation, and the right figure in the new implementation.}
\label{Fig:vectorStruc}
\end{figure}
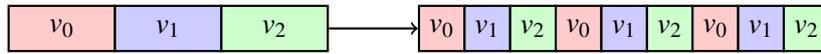

In concrete, we used pragmas such as \_\texttt{Pragma}("\texttt{unroll}"), \_\texttt{Pragma}("\texttt{vector aligned}"), and \_\texttt{Pragma}("\texttt{ivdep}") in front of these loops to signal compilers that they need to be vectorized.  In particular, these pragmas are applied to a \texttt{for}-loop of a pre-determined iteration length: \texttt{for( int i=0; i<num\_loop; i++)}.  Then, loops over rhs are written in terms of this basic loop, and so the number of rhs is required to be a multiple of \texttt{num\_loop}.  To accommodate our code to this loop strategy, we have prepared a new data structure for the bundle of vectors and rewritten all low-level routines involving loops over rhs to respect this new data structure.  Figure \ref{Fig:vectorStruc} schematically represents difference between old data structure and new data structure.  
As shown in the figure, in the old implementation, rhs are ordered so that they are placed one after the other.  In the new implementation, however, a given entry from all vectors are gathered and placed together, and the chunk of the next entry is put consecutively to the chunk associated with the previous entry so that the vector index is now the fastest running index.  

\begin{figure}[hbt]
\begin{subfigure}{.5\textwidth}
  \includegraphics[width=\linewidth]{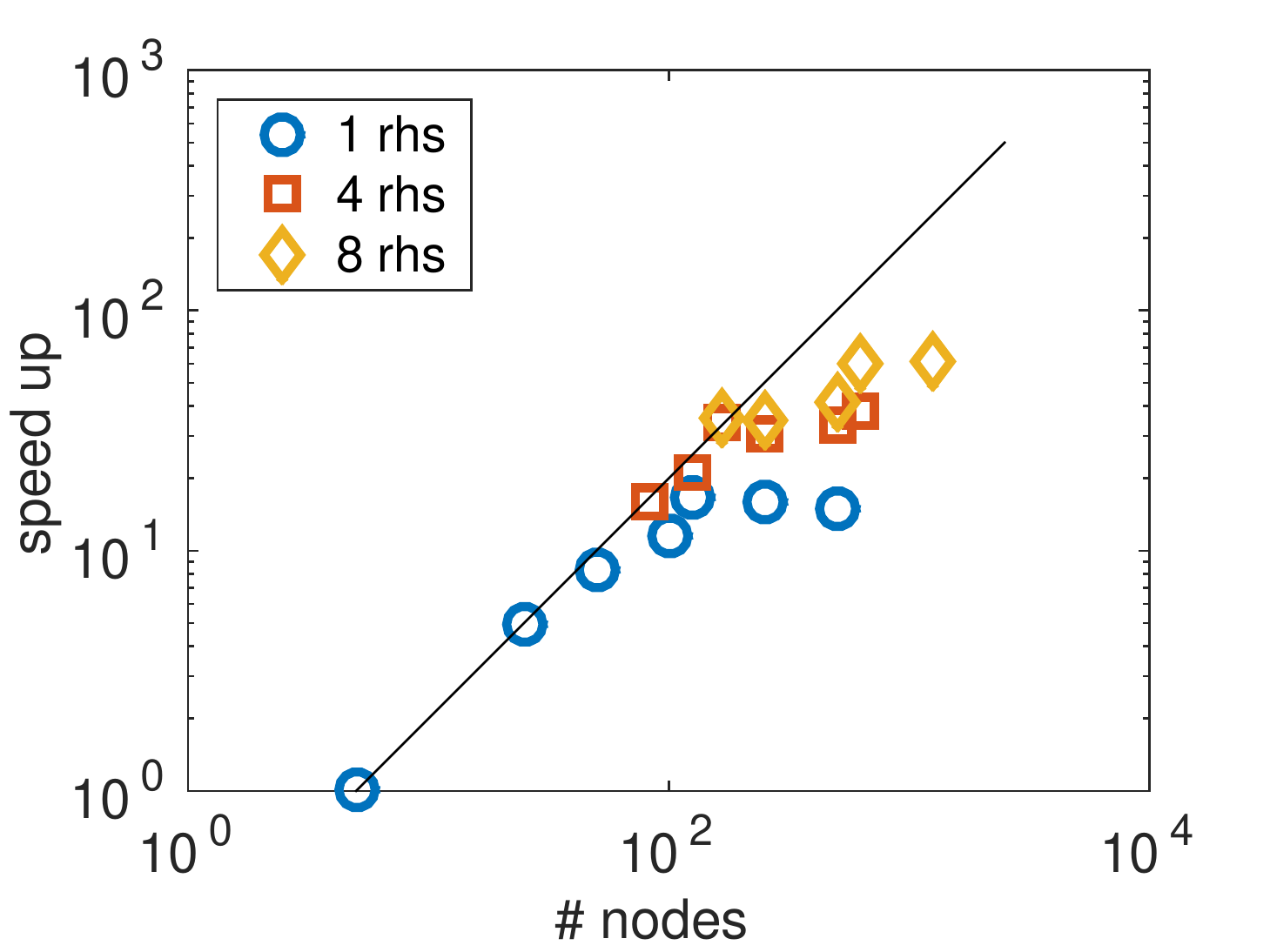}
  \caption{Relative speedup for time at the top levels}
  \label{Fig:scalingDDalphaAMG_mutipleRHS_total}
\end{subfigure}
\begin{subfigure}{.5\textwidth}
  \includegraphics[width=\textwidth]{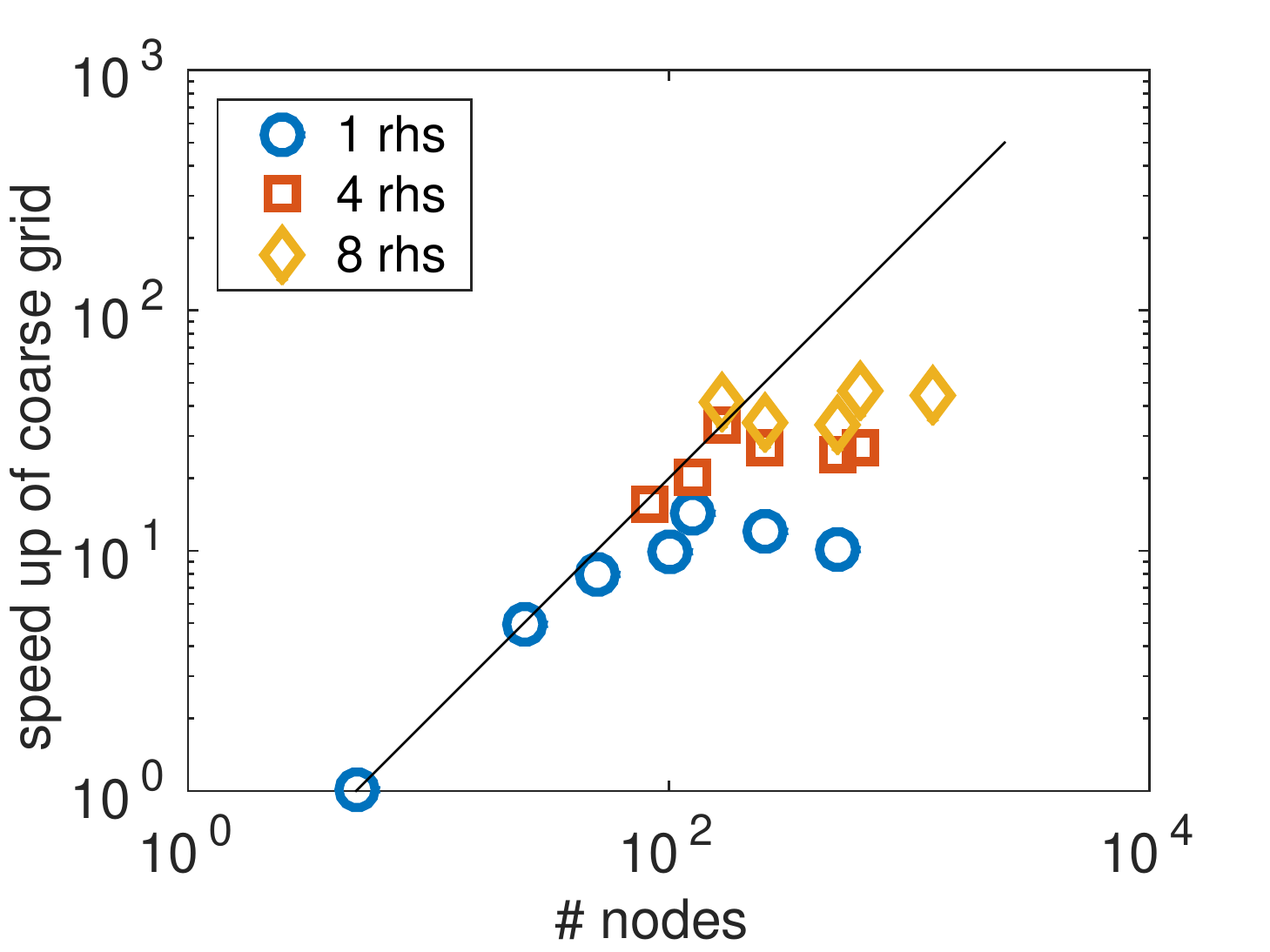}
  \caption{Relative speedup for time at coarse levels}
  \label{Fig:scalingDDalphaAMG_mutipleRHS_coarse}
\end{subfigure}
\caption{Comparison of scaling behavior with different number of r.h.s. as a function of the number of nodes used for simulation. }
\label{Fig:scalingDDalphaAMG_mutipleRHS}
\end{figure}

Using Intel\textsuperscript{\tiny\textregistered} Advisor for profiling, we found that the compiler automatically shifted vectorization from 128 bits to 256 bits for our new implementation of multiple rhs, as is shown in Table \ref{T:vectorization}.
\begin{table}[hbt]
\centering
\begin{tabular}{l | c | c | c | c | c | c }
Num. R.H.S. & \multicolumn{2}{c}{1 rhs} & \multicolumn{2}{c}{4 rhs} & \multicolumn{2}{c}{8 rhs}\\
Instruction Mix & SP Flops & DP Flops &  SP Flops & DP Flops & SP Flops & DP Flops  \\
\hline \hline
128-bit & 95.26\% & 86.59\% & 23.41\%  & 4.99\% & 24.92\% & 3.60\%\\ 
256-bit & 2.58\% & 1.26\%& 60.68\% & 78.13\% & 74.02\% & 94.76\%\\
Total & \multicolumn{2}{c}{97.26\%} & \multicolumn{2}{c}{84.03\%}  & \multicolumn{2}{c}{98.81\%} 
\end{tabular}
\caption{Vectorization Reports}
\label{T:vectorization}
\end{table}
It also has added benefit of reducing data loading time for the matrix, as the data need to be loaded only once for all rhs. Moreover, the scaling problem with DD$\alpha$AMG is mitigated via simultaneous inversion of multiple rhs.  Figure \ref{Fig:scalingDDalphaAMG_mutipleRHS} shows the scaling behavior of DD$\alpha$AMG with simultaneous inversion of multiple rhs.  As is shown in the figure, the breakdown of strong scaling is pushed away to higher parallelization. It now happens at around 512 nodes, widening the scalability window.  The right plot in the figure shows the scaling behavior at the coarsest level.  We see that the scaling behavior at the coarsest level is similar to the one at the top, which is based on the total computer time for inversion.  This suggests that the stagnation of speedup with the increasing number of cores can be traced to that at the coarse levels.  

\section{Block Solvers}
\label{Sec:BSolver}
Our implementation of simultaneous inversion of multiple rhs also works well with block solvers, which can accelerate convergence via extension of the search space by combining Krylov spaces for all rhs.  To enable various block Krylov solvers,  we linked our code to Fast Accurate Block Linear krylOv Solver (Fabulous), developed by Inria (France) \cite{ROBBE2006265,MORGAN2005222,doi:10.1137/140961912}.  Fabulous provides block solvers, namely block GMRES and GCR, with various features such as detection of inexact breakdown, deflated restarting, and incremental QR factorization.  This library also comes with two different orthogonalization schemes, Classical Gram-Schmit (CGS) and Modified Gram-Schmit (MGS) as well as their iterative variants each with a choice of blocking in orthogonalization.  Details can be found in the documentation at \url{https://gitlab.inria.fr/solverstack/fabulous}.  

With this library linked to DD$\alpha$AMG, it now provides block solvers at each level as an option for the solver for the coarse-grid problem in AMG in addition to non-block FGMRES.  The task is then to see if the use of block solvers leads to faster overall convergence at the top level.  For this, we needed to tune some of the parameters related to Fabulous as well as DD$\alpha$AMG to identify a parameter region  where DD$\alpha$AMG with block solvers performs better than without the block solvers. 

To set up the stage, we selected a three-level AMG with a FGMRES solver at the top level, as it is known to work best from our experience.  The goal residual at the top level was set to $10^{-10}$.  Among the parameters associated with DD$\alpha$AMG, the values of those parameters not related to coarse-grid inversion are set to be optimal values found in Ref.~\cite{Alexandrou:2016izb}.  
Among the various options for block solvers, inexact breakdown will not be considered in this study, as our implementation for multiple rhs always applies the Dirac matrix on at least $\texttt{num\_loop}$ many chunk of rhs.
As for orthogonalization scheme, it was found that CGS without iterative application generally performs better in our setup.  
In addition, GCR is the only flexible solver available in Fabulous so that we have considered only GCR and FGMRES as a middle-level solver.  Now, the parameters to be tuned are the choice of solvers at the middle and bottom and their target residuals.  In what follows, we first consider the solvers without deflation. 

The parameter tuning was conducted on Cyclone (Intel Xeon Gold 6248) at The Cyprus Institute using a lattice of size $48^3\times98$ at the physical point~\cite{ETM:2015ned}.
We have used $6$ nodes and $32$ cores from each node.  

\subsection{Tuning the residuals}
\begin{figure}[hbt]
\centering
\includegraphics[width=0.8\textwidth]{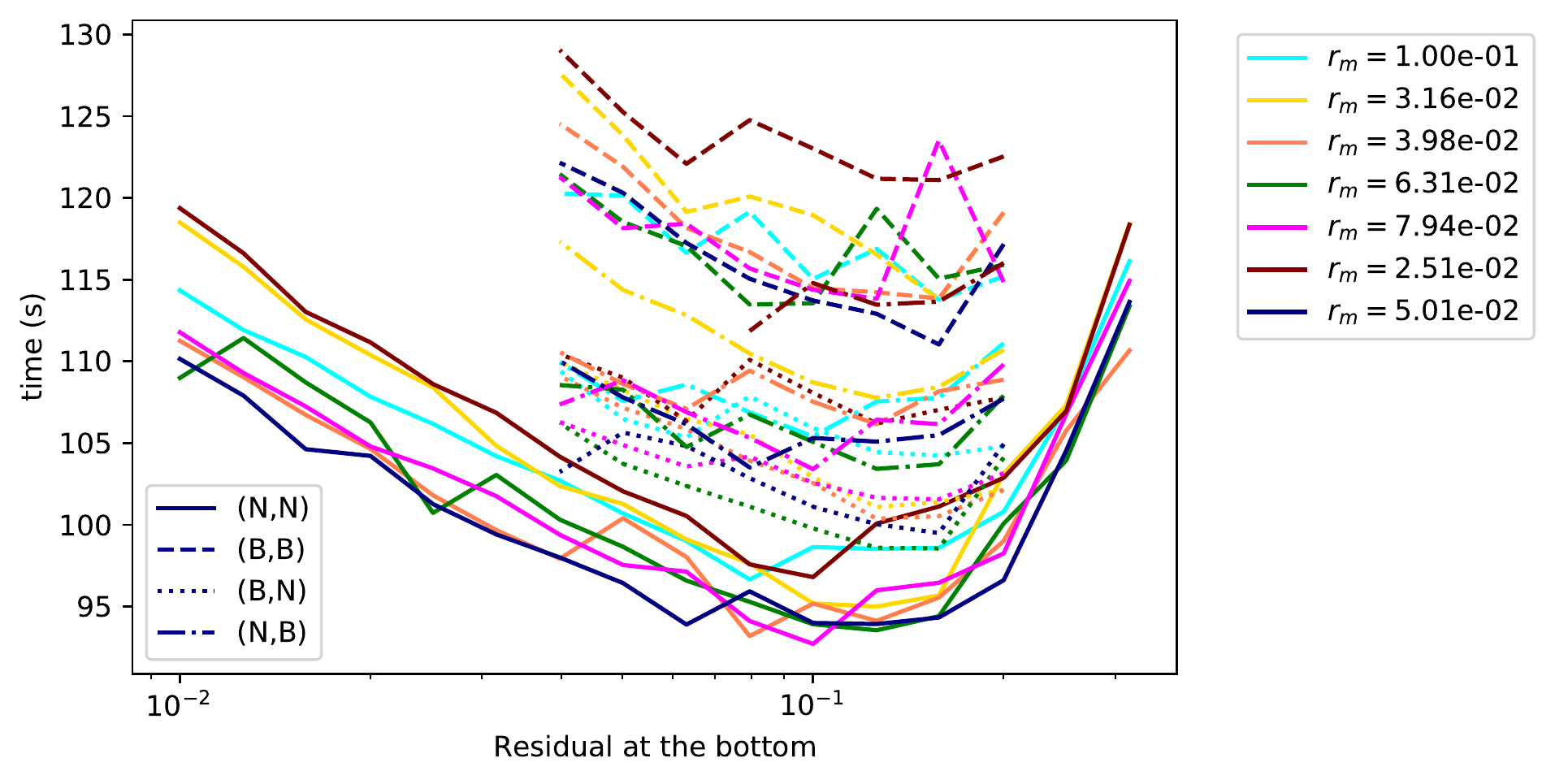}
\caption{Comparison of convergence time as a function of the bottom residual, $r_b$, for different middle residuals, $r_m$ between non-block AMG (solid lines) and mixed AMG.  The dashed line is associated with (Block, Block), the dotted line with (Block, Non-Block), and the dash-dotted line with (Non-Block, Block), respectively.}
\label{Fig:cresid_vs_solverTime}
\end{figure}

To find optimal parameter values with block solvers, we considered three cases: (Middle Solver, Bottom Solver) = (Block, Block), (Block, Non-Block), and (Non-Block, Block).  These cases were compared against the case (Non-Block, Non-Block).  For each case, we inverted 4 rhs with different bottom residuals for selected middle residuals.  Then, the results are compared with the (Non-Block, Non-Block) case, which is shown in Fig.~\ref{Fig:cresid_vs_solverTime}.  The slid lines are associated with purely non-block AMG, and other lines with mixed AMG.  The result indicates that purely non-block AMG leads to a faster convergence time.

\begin{figure}[hbt]
\begin{minipage}{0.5\textwidth}
\centering
\includegraphics[width=\linewidth]{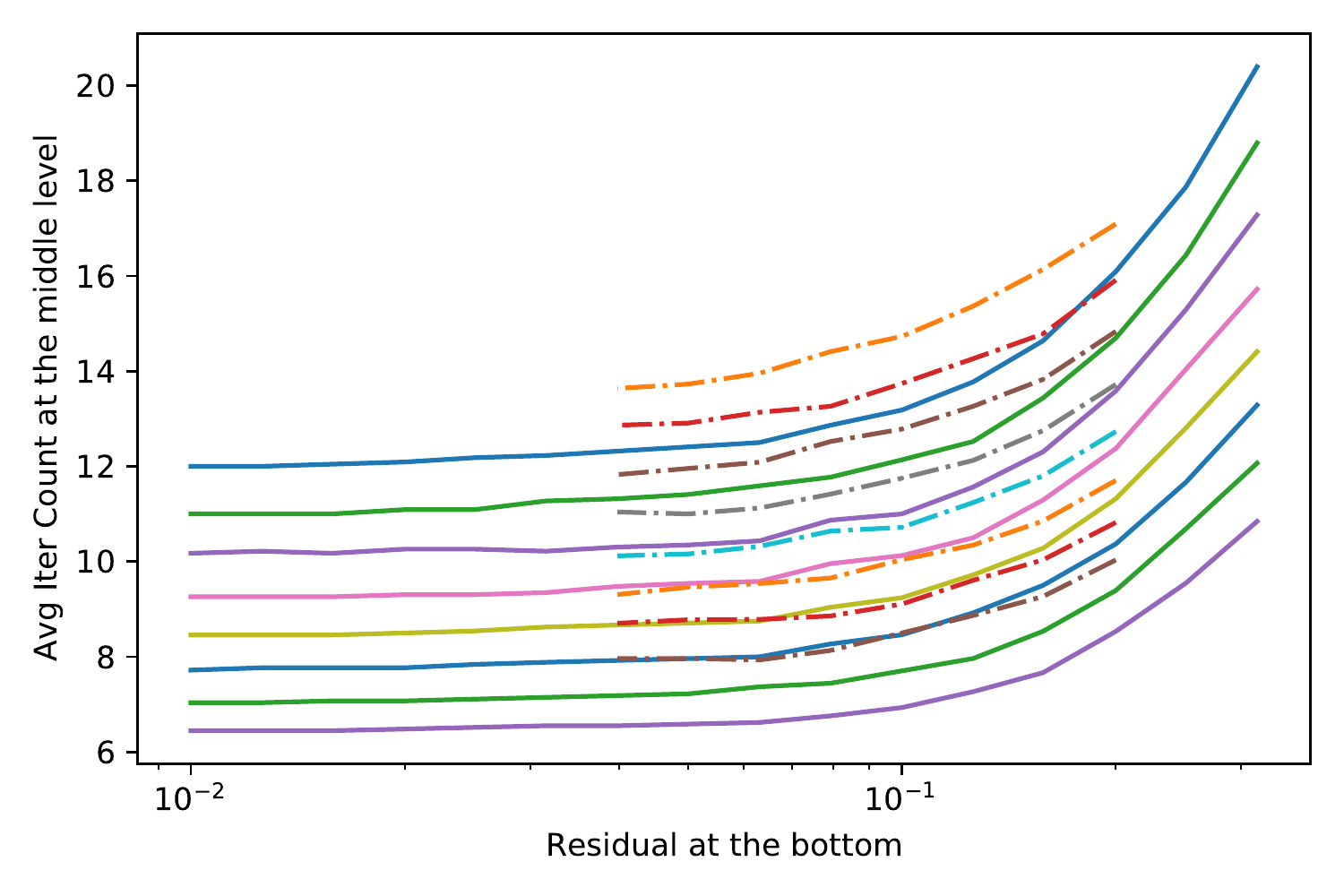}
\label{Fig:cresid_vs_itercount022_1}
\end{minipage}\hfil
\begin{minipage}{0.5\textwidth}
\centering
\includegraphics[width=\linewidth]{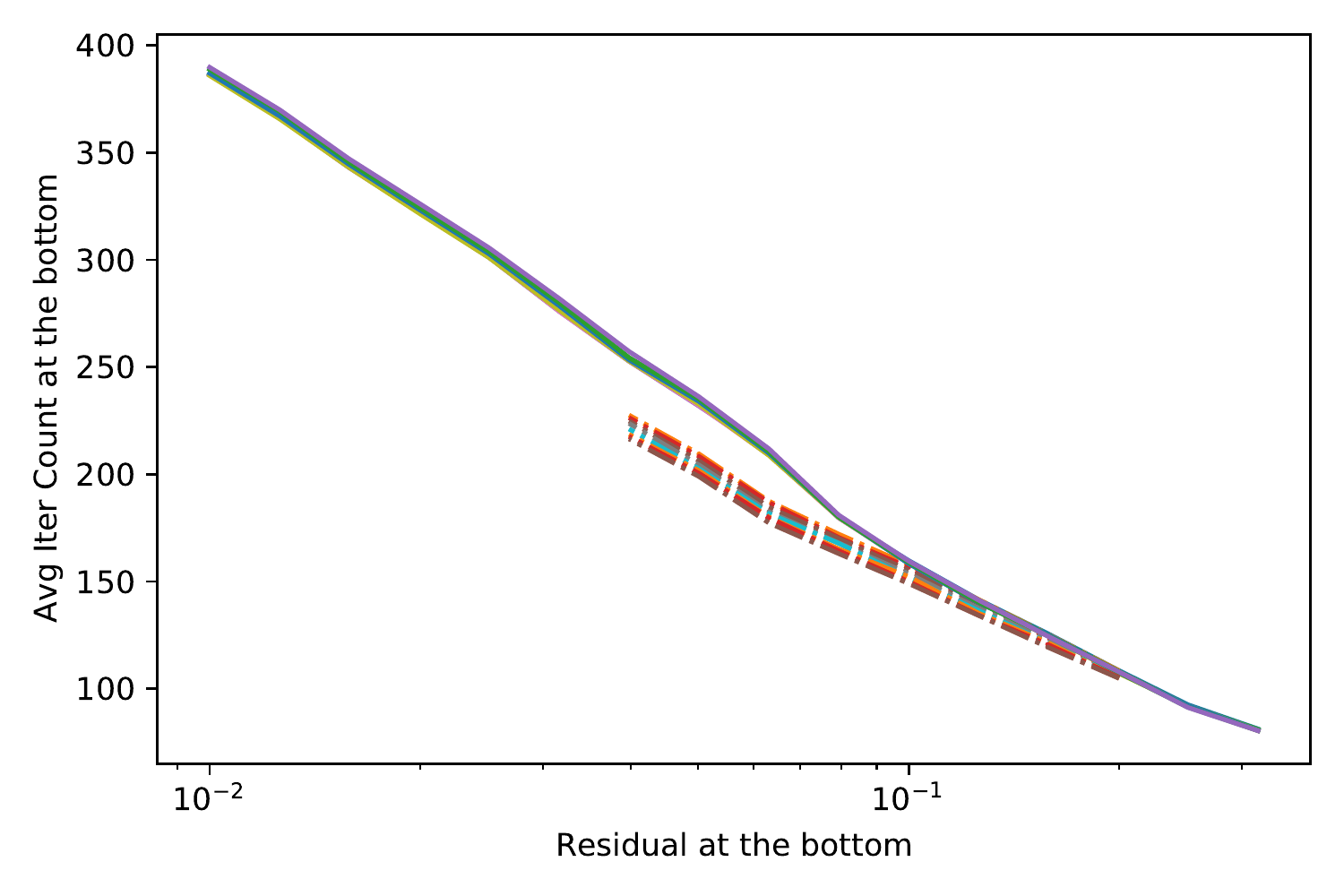}
\label{Fig:cresid_vs_itercount022_2}
\end{minipage}
\vskip -25pt
\caption{Comparison of average iteration counts between AMG with (Non-Block, Non-Block) in solid line and AMG with (Block, Block) in dashed line.}
\label{Fig:cresid_vs_itercount022}
\end{figure}

To see whether this is due to overhead in each iteration of block solvers or simply an indication of their ineffectiveness, we  compared the average iteration counts for convergence at each level over repeated calls to the coarse solver at the given level.  
Figure \ref{Fig:cresid_vs_itercount022} shows the comparison between non-block AMG and (Block, Block) AMG.  It reveals that the block solver as a middle solver requires more iterations than the non-block solver.
This is perhaps due to the small number of iterations required for convergence at the middle level.  At this level, the target residual does not need to be as small as the goal residual at the top level.  In our case, the goal residual at the top level is set to $1\times10^{-10}$, but the middle residual of around $0.5\sim1\times10^{-1}$ is small enough to ensure fast convergence at the top.  This is exhibited by Fig.~\ref{Fig:cresid_vs_solverTime} where the total convergence time takes the smallest values at $r_b \approx 1\times10^{-1}$ and $r_m \approx 7\times10^{-2}$.  
On the other hand, the right figure in Fig.~\ref{Fig:cresid_vs_itercount022} shows that  the average iteration counts with a block solver, indicated by dashed lines, is smaller than using a block solver below $r_b = 10^{-1}$.  Thus, block solvers can be effective when used as the bottom solver in the three-level multigrid.  
However, even in this case, AMG with block solvers requires more time to converge compared with AMG without block solvers.  To understand what constitutes the bottlenecks, we have used Score-P to trace and analyze performance of our code.  It turned out that the overhead originates from two major sources, namely reordering of vector layout, accounting for about 25\% of the total overhead, and global communication, which takes up about 75\%.  The first source comes from reordering of vector layout before and after each operator call within Fabulous.  This is necessary, as Fabulous uses a more conventional layout of multiple rhs where vectors are ordered one after the other, while in our code, the vector index is the fastest running index.  Also, Fabulous uses a user-provided matrix-vector-multiplication routine, and this routine is written for a new vector-running-fastest layout.  The other source is that the block-solvers increase the number of MPI-reduction calls significantly, by an order of magnitude. This is due to the enlarged search space, requiring more inner products.  This can be optimized by using non-blocking MPI-routines and pipelining. If available, they will lead to reduction of inversion time, to the extent that the block-solver becomes faster than the non-block version.
Note that due to the increase in computational workload in application of the coarsest operator to multiple rhs, some communication can be hidden in the piplelined version, which will directly minimize the overheads.

\begin{figure}[hbt]
\centering
\includegraphics[width=1\linewidth]{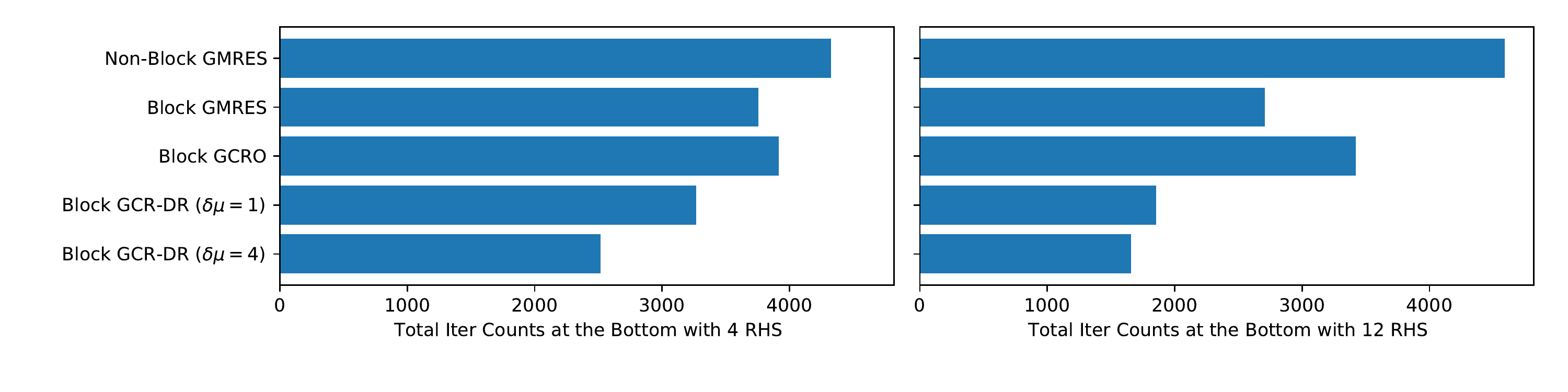}
\caption{Comparison of total iteration count at the bottom with non-block FGMRES as the middle solver at $r_m = 0.1$ and $r_b = 0.1$.  Here, $\delta\mu$ is a multiplicative factor in the definition of the Dirac matrix for twisted-mass fermions: $D_{TM} = D_W +i(\delta\mu)\mu\gamma_5$, provided as an parameter.}
\label{Fig:deflation4rhs}
\end{figure}

Lastly, we investigated effectiveness of deflation for the block Krylov solvers, which is available for BGRO-DR in the newest version of Fabulous, where DR stands for deflated restarting.  The results are summarized in Fig.~\ref{Fig:deflation4rhs}.   This shows a comparison of total iteration counts at the bottom with non-block bottom solver and block solvers with and without deflation.  
As can be seen, Fabulous solvers at the bottom with deflation are effective in reducing the total iteration counts.  They become more effective with the increasing number of rhs on account of a larger size of block Krylov spaces.  With 12 rhs, the number of the total iteration counts for Block BGCR-DR with $\delta\mu=1$ is around half of the iteration count in case of the non-block GMRES, but due to the overheads, the total inversion time is roughly the same as for the non-block variant.

\section{Outlook}
Multigrid methods, such as DD$\alpha$AMG, can significantly speed up inversion of linear systems including the one for the twisted-mass operator at physical quark masses.
While this has circumvented the problem of critical slowing down, the strong scalability  of multigrid methods is limited by the coarse grid size.

We show that simultaneous inversion of multiple right-hand sides can overcome this issue and widens the scalability window while reducing data loading time and improving portability and maintainability of the code at the same time.
Further improvements are possible by using block Krylov solvers at the coarsest level.  We tested various block Kylov solvers with DD$\alpha$AMG by linking it to an external library, Fabulous. 
This involved tuning some parameters, such as the residual on the coarse grids. In general, we found that block Krylov solvers can reduce iteration count by up to a factor 2 in comparison to the optimized native GMRES method, especially if we use BGCRO-DR. Note that BGCRO-DR can be used without an additional shift parameter on the coarsest grid in case of twisted mass fermions.

At the moment, the use of Fabulous comes with an additional overhead due to reordering of vectors in each DD$\alpha$AMG kernel call and due to blocking of global communication functions.  This overhead can be mitigated by using pipelined block-Krylov solvers and modifying inner-product routines, which could lead to improvements in inversion time by up to a factor 2.

\section{Acknowledgments}
This work was supported by computing time awarded on the Cyclone supercomputer of the High Performance Computing Facility of The Cyprus Institute under project p009.  This project has received funding under PRACE-6IP, Grant agreement ID: 823767, Project name: LyNcs. LyNcs is one of 10 applications supported by PRACE-6IP, WP8 ``Forward Looking Software Solutions''. S.Y., S.B.~and J.F.~have received funding under this project. The authors would also like to thank the other members of the LyNcs project for the stimulating collaboration, with special thanks to Luc Giraud and Matthieu Simonin from Inria, Bordeaux and Michele Martone from LRZ, Munich.

\bibliographystyle{JHEP}
\bibliography{refs}

\end{document}